\documentclass[aps,pra,reprint,superscriptaddress,showpacs]{revtex4-1} 
\usepackage{graphicx,epstopdf}
\usepackage{amsmath}  
\usepackage{ulem}
\usepackage{CJK}
\usepackage{subfigure}

\usepackage[raggedright]{titlesec}
\usepackage{titlesec}

\newcommand{\upcite}[1]{\textsuperscript{\textsuperscript{\cite{#1}}}}

\usepackage{hyperref}
\hypersetup{hypertex=true,
            colorlinks=true,
            linkcolor=blue,
            anchorcolor=blue,
            citecolor=red}

\begin{document}
	
\title{Homogeneous Linear Ion Crystal in a Hybrid Potential}

\author{Ming-shen Li}	
\thanks{These two authors contributed equally}
\affiliation{School of Physics and Astronomy, Sun Yat-Sen University, Zhuhai, 519082, China}
\author{Yang Liu}
\thanks{These two authors contributed equally}
\affiliation{School of Physics and Astronomy, Sun Yat-Sen University, Zhuhai, 519082, China}
\affiliation{Center of Quantum Information Technology, Shenzhen Research Institute of Sun Yat-sen University, Nanshan Shenzhen 518087, China}
\author{Xin-Xin Rao}	
\affiliation{School of Physics and Astronomy, Sun Yat-Sen University, Zhuhai, 519082, China}
\author{Peng-Fei Lu}
\affiliation{School of Physics and Astronomy, Sun Yat-Sen University, Zhuhai, 519082, China}
\author{Zhao Wang}
\affiliation{School of Physics and Astronomy, Sun Yat-Sen University, Zhuhai, 519082, China}
\author{Feng Zhu}
\email{zhufeng25@mail.sysu.edu.cn}
\affiliation{School of Physics and Astronomy, Sun Yat-Sen University, Zhuhai, 519082, China}
\affiliation{Center of Quantum Information Technology, Shenzhen Research Institute of Sun Yat-sen University, Nanshan Shenzhen 518087, China}
\author{Le Luo}
\email{luole5@mail.sysu.edu.cn}
\affiliation{School of Physics and Astronomy, Sun Yat-Sen University, Zhuhai, 519082, China}
\affiliation{Center of Quantum Information Technology, Shenzhen Research Institute of Sun Yat-sen University, Nanshan Shenzhen 518087, China}
\affiliation{School of Physics and Astronomy, Sun Yat-sen University, Zhuhai 519082, China}	
	
\date{\today}	

\begin{abstract}
We investigate the properties of a linear ion crystal in a combination of quadratic and quartic potentials. Both the discrete and the continuous model are employed to explore the homogeneity of a linear ion crystal by controlling the proportional parameter between the quadratic and quartic components. It is found that a uniform ion distribution in such a hybrid potential can be made larger than that in a purely quadratic or quartic potential. The zigzag transition is also investigated in the hybrid potential. Its critical condition and phase diagram are determined numerically, which agrees well with previous theoretical and experimental results. This paves the way for experimental investigation of phase transition in a large linear coulomb crystal.
\end{abstract}


\maketitle

\section{Introduction}

Quantum computers have the potentials to tackle many problems hard to be solved or even unsolvable using classical computers.\upcite{kaye2007introduction, james1998quantum} 
Owing to the long-lived internal qubit coherence, the ion trap becomes one of the most attractive candidates for the realization of a quantum computer.
Today's prototype quantum computers containing tens of qubits have been demonstrated in several systems including ion traps. To realize the promise of quantum computing, intensive efforts had been paid to construct quantum computer with a large scale of ion qubits.\upcite{cirac1995quantum,sorensen1999quantum,milburn2000ion} Different types of ion traps had been investigated, such as the linear ion trap designed to trap a long ion chain with even space,\upcite{johanning2016isospaced,xie2017creating} the ion trap on chips \upcite{graham2014system} to facilitate ion transportation in different areas, the collection of small-scale ion traps integrated by optical methods,\upcite{blain2006microfabricated} etc.

A linear chain of trapped ions in a harmonic potential has been demonstrated as a quantum computer of tens of qubits. \upcite{dubin1997minimum,bastin2017ion,johanning2016isospaced,ohira2020confinement} However, in the harmonic potential it is difficult to address single ion due to the unequal spacing. The zigzag transition will occur as the number of ions increases. \upcite{schiffer1993phase} Therefore, it is requisite to realize the uniform distribution of a long ion chain for a large-scale quantum information system. The uniform ion chain is critical for the single and multiple qubit processing and gate operation. It would also provide an ideal test-bed for the study of many-body effects such as topological phase transition. \upcite{gutierrez2020defect} In addition, under the same radial confinement, the evenly spaced linear ion chain can accommodate more ions than in the harmonic potential before the zigzag transition occurs. \upcite{lin2009large}

To achieve a uniform distribution of a long chain, ion distributions in anharmonic potentials have been studied.\upcite{bastin2017ion} The results showed that the distribution of the ions in an anharmonic potential is more uniform at the center than that in a harmonic potential, but it changes quickly at both edges. Additionally, the combination of the harmonic and anharmonic potentials was proposed as the trapping potential,\upcite{lin2009large} in order to scale up the qubits and to operate the entangling gates more efficiently. Recently, using anharmonic axial potentials, 100 ions in a linear configuration have been realized and up to 44 ions have nearly equidistant spacings \upcite{pagano2018cryogenic} Prospects for even larger system sizes look quite optimistic in the near future. 

In this paper, we present a homogenous crystal with large number of ions trapped in a hybrid potential, consisting of harmonic and anharmonic potentials with a dimensionless ratio. First, we use the discrete model to calculate the requisites for making a homogenious ion chain by using the hybrid potential. Then, we test our calculations using a universal variance criterion. To compare with the discrete model, we further apply variational method to the hybrid potential by using the continuous model. Almost identical conditions are obtained, which confirms the accuracy of our calculations and the demanding of a well-designed hybrid potential in order for a large homogenious ion chain. At the end, we discuss the zigzag transition in the hybrid potential.

\section{Discrete model}

The discrete model has been applied to the linear ion chain in a purely quadratic(i.e. harmonic) or quartic(i.e. anharmonic) potential. \upcite{dubin1997minimum, james1998quantum, bastin2017ion} The potential energy of the system in a quadratic potential can be described as: \upcite{dubin1997minimum, james1998quantum}
\begin{eqnarray}
e=\sum_{i=1}^{N}\dfrac{1}{2}m\omega^2x_{i}^2+\sum_{i,j=1;i<j}^{N}\dfrac{q^2}{\left|x_{i}-x_{j}\right|}
\end{eqnarray}
where $m$ and $q$ are the mass and charge of an ion, $\omega$ is the axial trap frequency, $x_{i}$ is the axial position of each ion, $N$ is the total number of ions. When replacing $m\omega^2x_{i}^2/2$ with $ax_{i}^4/4$, the confinement becomes a purely quartic potential, in which $a$ is the coefficient. 

Here, we first use the discrete model to calculate the ground state distribution of the ion chain in the hybrid potential. The energy of the system can be written in the following form:\upcite{lin2009large}
\begin{eqnarray}\label{dis_E}
e=\sum_{i=1}^{N}\frac{\alpha_{4}}{4}x^4_{i}-\sum_{i=1}^{N}\frac{\alpha_{2}}{2}x^2_{i}+\sum_{i,j=1,i<j}^{N}\frac{q^2}{|x_{i}-x_{j}|}
\end{eqnarray}
This equation can be rescaled by $z_{i}=x_{i}/l$ with  $x_{i}$ with a length scale $l^5= q^2/\alpha_{4}$. Through a dimensionless processing, Eq.(\ref{dis_E}) can be rewritten as:
\begin{eqnarray}
E=\sum_{i=1}^{N}\frac{1}{4}z^4_{i}-\sum_{i=1}^{N}\frac{D}{2}z^2_{i}+\sum_{i,j=1,i<j}^{N}\frac{1}{|z_{i}-z_{j}|},
\end{eqnarray}
using a dimensionless ratio $D=|\alpha_{2}/q^2|^{2/5}(\alpha_{2}/\alpha_{4})^{3/5}$, which is introduced to characterize the strength of the quadratic potential relative to the quartic potential. 

\begin{figure}[htb]\vspace*{-8pt}
	\includegraphics[width=3.3in]{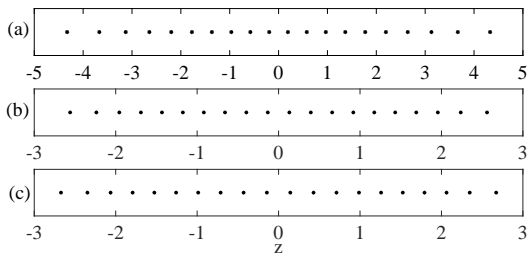}
	\caption{\label{figPosition20}Positions of $20$ ions in (a) quadratic, (b) quartic and (c) the hybrid potentials. Notice that the ions in the hybrid potential are more uniformly distributed. In these plots, the position z has been rescaled to be dimensionless. }
	\label{Position}
\end{figure}

The ground state positions of the ions can be determined by the following conditions
\begin{eqnarray}
\frac{\partial E}{\partial {z}_{i}}=0 \label{pEmin}
\end{eqnarray}
By solving totally $ N $ nonlinear differential equations, we obtain the equilibrium position of each ion. The positions of $20$ ions in different potentials are compared in Fig.\ref{Position}. The ions in the hybrid potential are much more evenly distributed across the chain compared to the quadratic or quartic potential.

\begin{figure}[htbp]
 \includegraphics[width=3.3in]{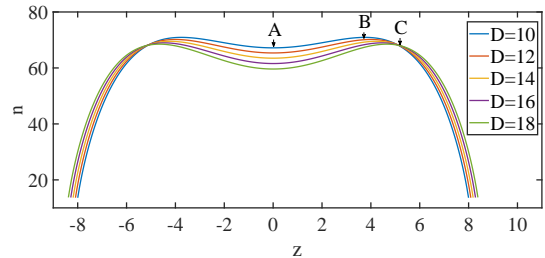}
 \caption{\label{figdensigy} The ground state ion density $n$ in hybrid potentials for different $ D $ in discrete model, when the total number of ions is $N=1000$ . In the middle part of the ion chain (i.e. around $z=0$), the density decreases with the increase of $D$. A represents the midpoint of the ion chain, B represents the point where the density reaches the maximum, and C resprents the point where the density is the same as point A. Both $n$ and $z$ are rescaled to be dimensionless.}
 \label{N1000}
\end{figure}

By defining the ion density as follows:
\begin{eqnarray}
n(\tilde{z_{i}})=\frac{1}{|z_{i+1}-z_{i}|},
\end{eqnarray}
where ${z_{i}}$ is the coordinate of the $i$th ion, $\tilde{z_{i}}$ is the midpoint between ${z_{i}}$ and ${z_{i+1}}$, the resultant ion density distribution is shown in Fig.\ref{N1000} for different $D$ parameters when $N=1000$. It can be seen from Fig.\ref{N1000} that the density in the middle part of the ion chain decreased with the increase of the $D$ parameters. This is because when the parameter $ D $ increases, the extreme point of the hybrid potential deviates farther from the center, resulting in an increase of the ion spacing at the center.

\begin{figure}[htbp]
	\centering
	\subfigure{\includegraphics[width=3.3in]{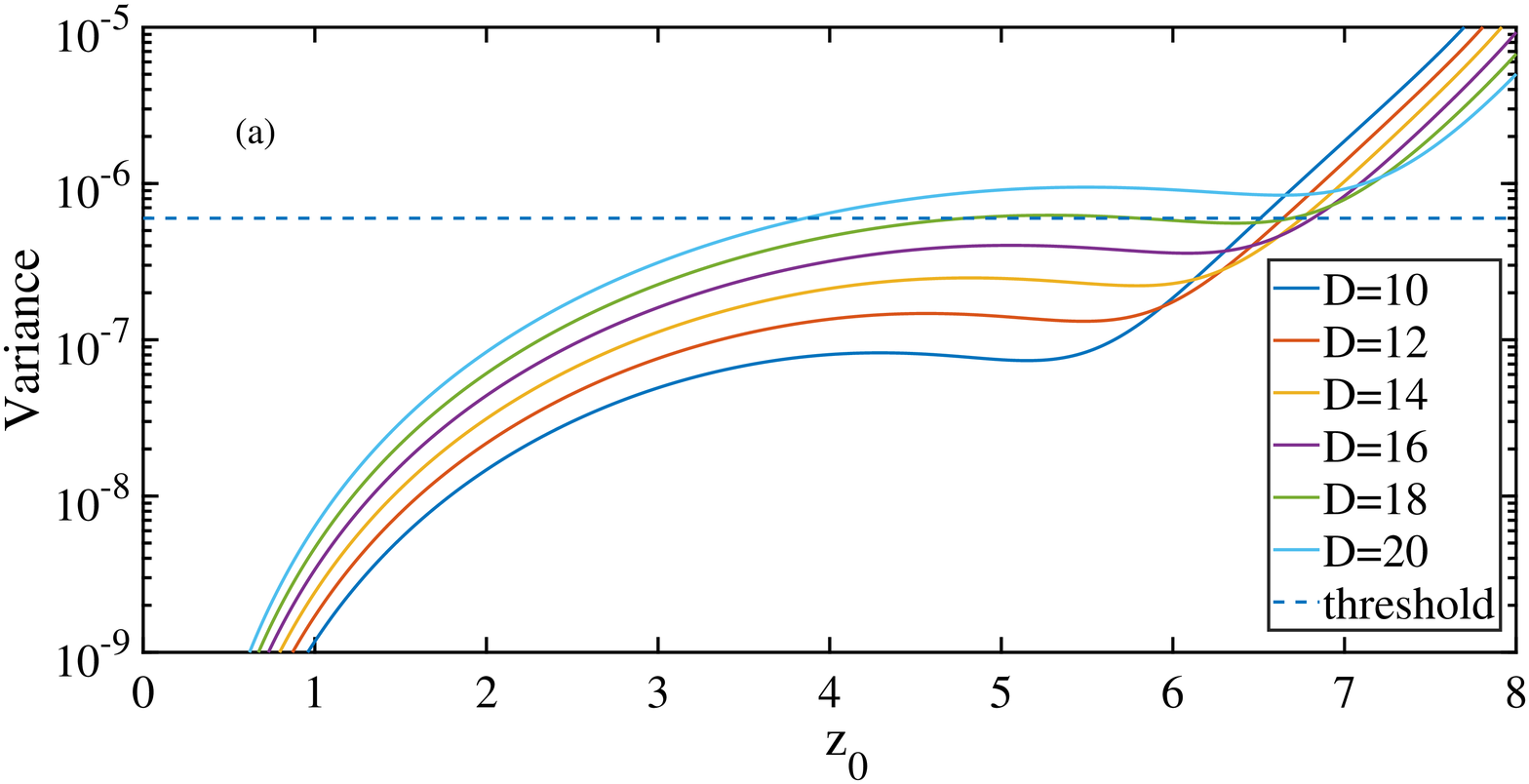}}\vspace{-10pt}
	\subfigure{\includegraphics[width=3.3in]{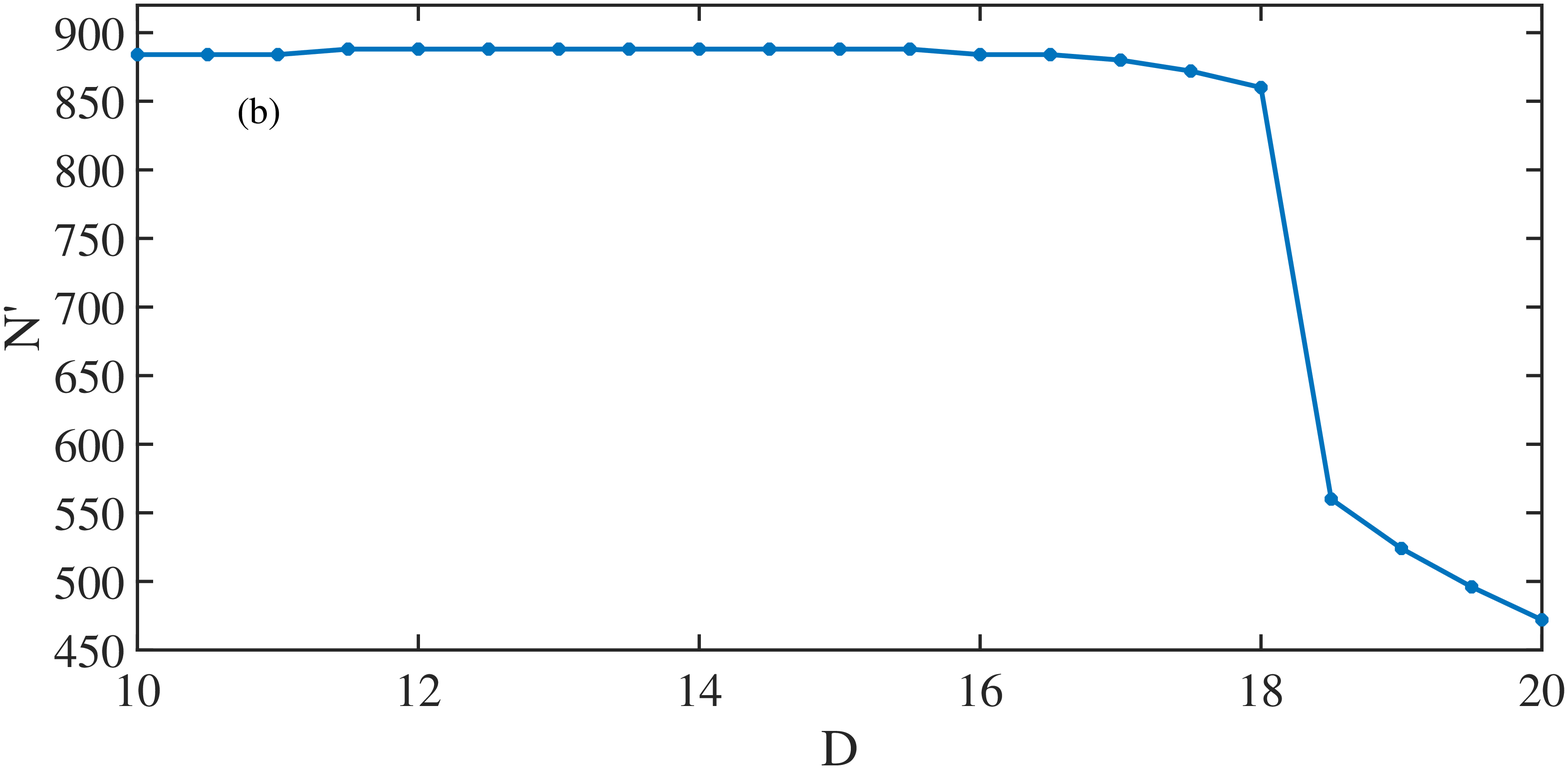}}\vspace{-10pt}
	\caption{\label{figdensigy} (a) Variance criterion in discrete model, as a function of $z_0$ with different $D$ parameters for $N=1000$. The dashed line represents the threshold $\eta = 6\times 10^{-7}$, and we choose the right intersection point between the dashed line and the curve for determining the optimal length of the homogenious ion chain for a $D$ parameter, which in turn corresponds to an optimal ion number $N^{\prime}$ for this chain. Here, $z_{0}$ are rescaled to be dimensionless. (b) The optimal number of ions as a function of $D$ parameter for the same $\eta$. When $ D >18 $, a sharp reduce in $N^{\prime}$ appears. This is due to a smaller $z_0$ with a larger $D$ in (a). }
	\label{discrete}
\end{figure}

In order to find an appropriate combination of the quadratic and quartic potentials to maximize the uniform distribution of the ions, the variance $ s(z)=\frac{1}{N}\sum_{i=1}^{N}(\Delta z_{i} - \overline{\Delta z_{i}})^2$ of $\Delta z$ in $[0,z_0$] is introduced. Fig.\ref{discrete}(a) shows the variance as a function of the position $z_0$. The optimal distribution is set by the criterion of $s(z) \le \eta$. The optimization of $D$ parameter is shown for $\eta = 6\times 10^{-7}$ and $ N=1000 $. We define the optimal number of ions that satisfy the optimization criterion to be $N^{\prime} $. As shown in Fig.\ref{discrete}(b), the optimal number of ions fluctuate little for $ 10< D <18 $ since these curves intersect with the threshold in a small range. The optimal ion numbers of $N^{\prime}$ have a maximum of $ 888 $ when $ D\in[11.5,15.5] $. Under the same conditions, $N^{\prime}$ is $ 396 $ or $ 844 $ in the quadratic or the quartic potential respectively from our caculation, which is significantly lower than the optimal ion number in the hybrid potential. 

\section{Continuous model}

Although the discrete model is effective in obtaining requsites for a homogenious distribution of an ion chain, it is very time-consuming especially when the number of ions is hundreds or thousands.  Thus, in order to solve the properties of the ion chain more efficiently and also to confirm our calculations with discrete model, we implement continuous model based on Dubin's variational method \upcite{dubin1997minimum}, which revolves around the assumption of the local density approximation, \upcite{hohenberg1964inhomogeneous} where the distribution of each ion in an ion chain is approximated by a continuous density function $n(z)$. Time consumption using this method does not relate to the scale of the total ion number $N$, thus making the calculation much faster and more flexible.

The first term in the integral is the energy due to the quadratic trapped potential. When replacing ${\frac{1}{2}z^2}$ with ${\frac{1}{4}z^4}$ it becomes the energy due to the quartic potential. The other term is the Coulomb energy of the ion chain and $\gamma\approx 0.57721 $ is Euler’s constant. The principle of minimum energy is used to minimize Eq.(\ref{eq:min_energy_integral}), and the density function $n(z)$ can be obtained.The variational method was extended to an quartic potential recently.\upcite{bastin2017ion} 

\begin{figure}[htbp] 
    \includegraphics[width=3.3in]{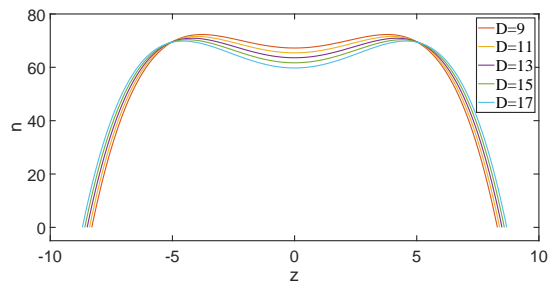}\vspace{-10pt}
	\caption{\label{figdensigy} The ground state density functions calculated by the variational method when $N=1000$. The shapes of the curves are identical to the corresponding ones calculated by the discrete model in Fig.\ref{N1000}. Both $n$ and $z$ are rescaled to be dimensionless.}
	\label{VN1000}
\end{figure}

According to the local density approximation, the density function of the ions in a hybrid potential can be expressed as $n(z)=Az^4+Bz^2+C$.\upcite{dubin1997minimum,bastin2017ion} By substituting the boundary condition and the normalization condition, it becomes:
\begin{equation}	
n(z)=-\frac{5(-2C+6CL-3N)}{2L^4 (-5+3L)}z^4+\frac{3(8CL-5N)}{2L^2 (-5+3L)}z^2+C, \label{vpoly}
\end{equation}
which is shown in Fig.\ref{VN1000}. It is worth to notice that both $L$ and $C$ are functions of $N$ and $D$, since the distribution of the ions is determined by $N$ and $D$ according to the principle of the minimum energy. It is found that these parameters have extremely complicated representations with $N$ and $D$ as variables. However, we find simple expressions for these two parameters, by using numerical methods such as gradient search algorithm, such that the potential energy of of the system \upcite{bastin2017ion}
\begin{equation}
\begin{aligned}
	E[n]=&\int_{-\infty}^{\infty}dz\{\frac{1}{2}z^2n(z)+\gamma n(z)^2 \\
	&-\frac{1}{2}n(z)\int_{0}^{\infty}dy\ln{[yn(z)]}\frac{d}{dy}[n(z-y)+n(z+y)]\}\label{eq:min_energy_integral}
\end{aligned}
\end{equation}
reaches its minimum. Notice here, a quartic term of $\frac{1}{4}z^4$ representing anharmonic potential shall be added into the integral while using this expression. When $N$ is in the order of hundreds or thousands, the $C$ and $L$ can be formulated as:
\begin{eqnarray} \label{pa}
	&C=0.26N^{0.80}-0.36D^{1.23}+3.80\\
	&L=0.70N^{0.30}+0.04D^{1.13}+2.16
\end{eqnarray}
Consequently, the density function is determined, as shown in Fig.\ref{VN1000}. 

\begin{figure}[htbp]
	\centering
	\subfigure{\includegraphics[width=3.3in]{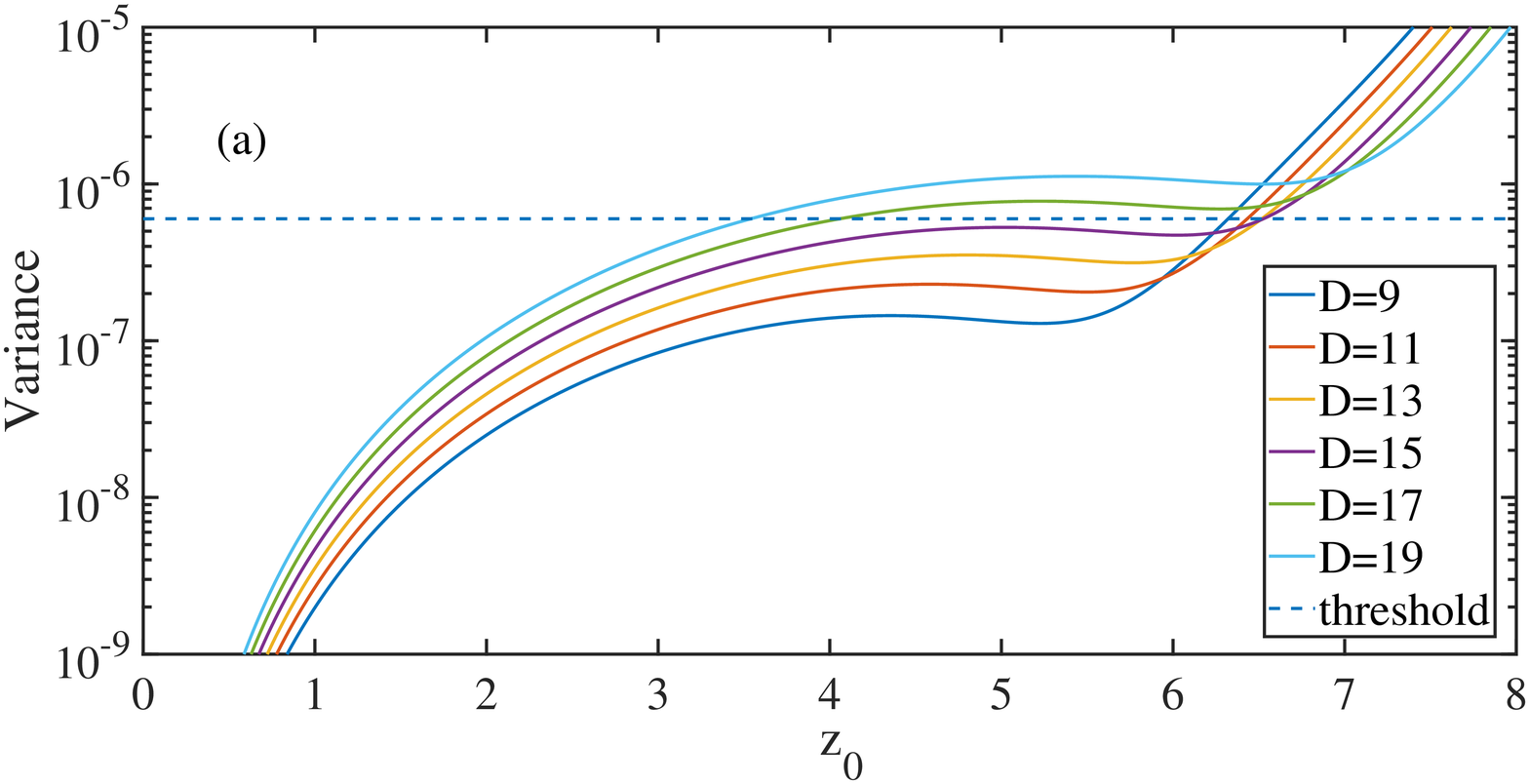}}\vspace{-10pt}
	\subfigure{\includegraphics[width=3.3in]{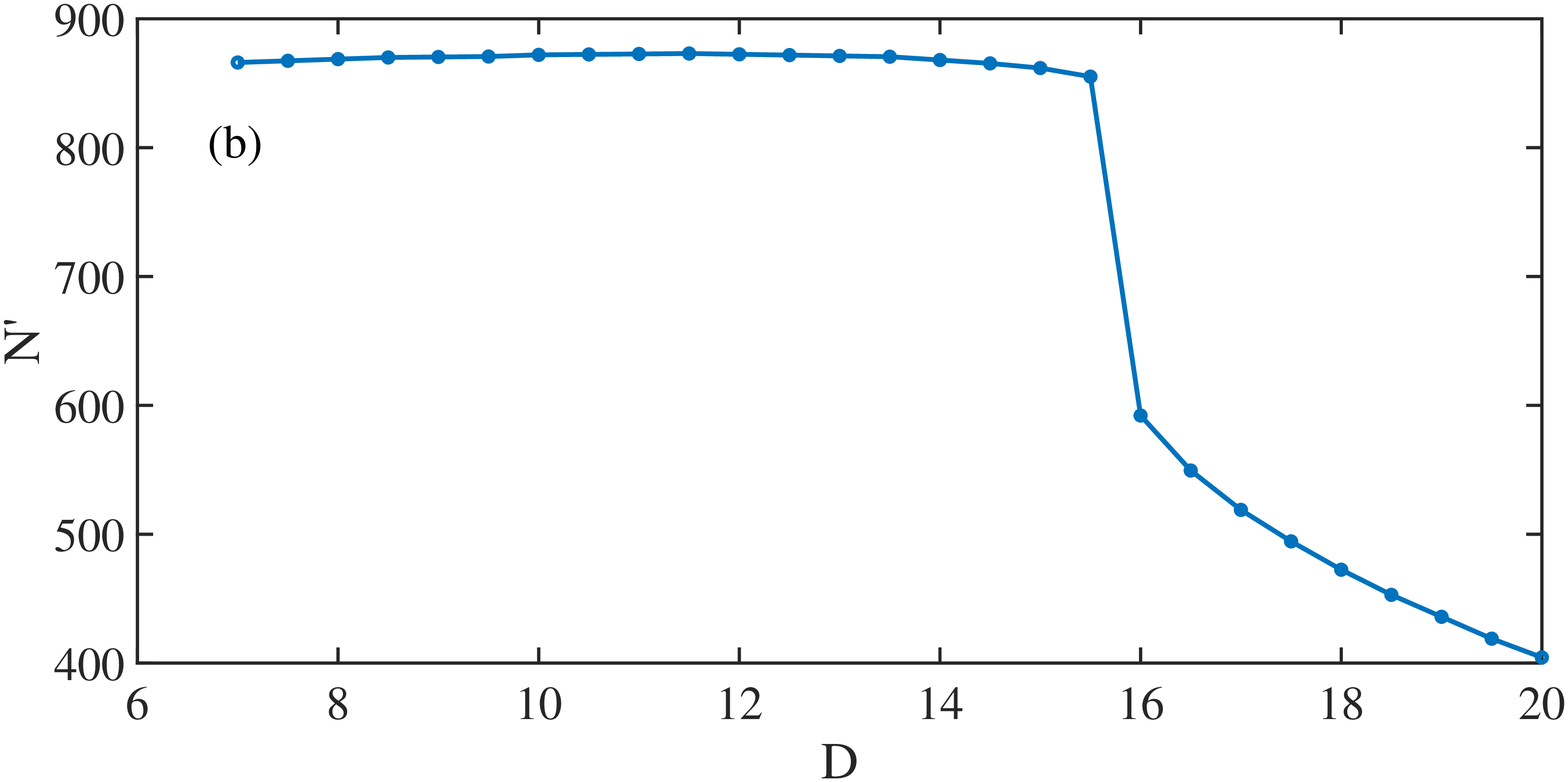}}\vspace{-10pt}
	\caption{\label{figdensigy}(a) Variance criterion as a function of $z_0$ with different $D$ parameters for $N=1000$. The dashed line represents the threshold of $\eta = 6\times 10^{-7}$, and the intersections between the dashed line and the curves determine the available lengths of the ion chains for different $D$ parameters, which can be converted into ion numbers by the density function. (b) The optimal number of ions as a function of $D$ parameters with the $\eta = 6\times 10^{-7}$. }
	\label{continuous}
\end{figure}

Using global density function \label{vpoly}, it is convenient to analyze the whole distribution uniformity to find the number of the evenly distributed ions. In order to demonstrate the characteristic of the general distribution, we define the variance criterion as:
\begin{eqnarray}
	s(z_0)=\frac{1}{z_0}\int_{0}^{z_0}[\Delta(z)-\frac{1}{z_0}\int_{0}^{z_0}\Delta(z)dz]^2 dz
\end{eqnarray}
where $\Delta(z)=\frac{1}{n(z)}$ represents the distance between the adjacent ions. The variance criterion is a function of position $z_0$, as shown in Fig.\ref{continuous}(a).

With the continuous model, we have $N(z_{0})=\int_{0}^{z_{0}}n(z)dz$, where $N(z_{0})$ represents the particle number in the range of $ [0, z_{0}]$. Once we determine the appropriate $z_0$, we get the number of particles. The variance criterion of the density function can have different requirements restricted by the design of the experiment. Fig.\ref{continuous}(b) show, the number of ions with homogenious distribution depends heavily on $ D $ parameter, same as the trend shown in Fig.\ref{discrete}(b). If the variance criterion $s\le 6\times 10^{-7}$, we have $N^{\prime}=872$ when $D\in[10,12]$, while $N^{\prime}=796$ for quartic potential and $380$ for quadratic potential from our caculation. We also notice there are small differences between critical $D$ value between continuous model and discrete model, which is also shown in \cite{bastin2017ion} and is due to the differences introduced by the integration by the variational method. 

\section{Zigzag phase transition}

During the discussion above, we assume the motion in the radial direction is frozen out by a tight radial potential which is much stronger than the axial potential, so we ignore the effect of radial potential. However, when the radial confinement is not that strong, it is energetically favorable for the ions to be displaced in the radial direction, and the zigzag transition occurs,\upcite{dubin1993theory, nigmatullin2016formation, pedregosa2020defect} which is proven to be a second-order phase transition. With the results of the continuous model, we analyze the zigzag transition under the condition that $x$-potential is much weaker than $ y $-potential, so the $ y $-potential can be ignored. We use $\beta$ as the strength of the $x$-potential relative to $z$-potential, and the anisotropy of the trap can be approximated by the $\alpha (z\to 0) = \frac{\omega_{x}^2}{\omega_{z}^2} (z\to 0) \approx \beta^{2}/D$, because it is always the case the phase transition first occurs at the center of the ion chain($z=0$). The potential energy of the system is then:
\begin{equation}
	\begin{aligned}
		e=&\sum_{i=1}^{N}(\frac{\beta ^2}{2}x_i^2+\frac{1}{4}z^4_{i}-\frac{D}{2}z_{i}^2)\\
		&+\sum_{i,j=1;i<j}^{N}\frac{1}{|(x_{i}-x_{j})^2+(z_{i}-z_{j})^2|^{\frac{1}{2}}}
	\end{aligned}
\end{equation}

The motion of the ions can be described by the Lagrangian,\upcite{james1998quantum,enzer2000observation,kielpinski2000sympathetic}where the Hessian matrix can be expressed as:\upcite{bastin2017ion}
\begin{eqnarray}
A_{mn}=\left \{\begin{array}{c}(\beta^2-\sum_{p\not =n} \dfrac{1}{|z_p-z_n|^3}),\  \ \ \ m=n,\\ \\\ \ \ \ \ \ \ \ \ \ \ \ \dfrac{1}{|z_m-z_n|^3},\ \ \ \ \ \ \ \ \  m\not = n\end{array}\right .
\end{eqnarray}
The eigenvectors and eigenvalues of the real, symmetric, positive-definite matrix $A_{mn}$ define the normal modes of oscillation of the ions along the $z$ direction. The eigenvectors $b_{m}^{(p)}$ are defined by $\sum_{m=1}^{N}A_{mn}b_{m}^{(p)} = \mu_{p}b_{n}^{(p)}$, where eigenvalue $\mu_{p}\geq 0$ and $p = 1,\cdots, N$ are the mode index. The eigenvectors are normalized and enumerated in order of increasing eigenvalue. 

\begin{figure}[htbp]
	\centering
	\includegraphics[width=3.3in]{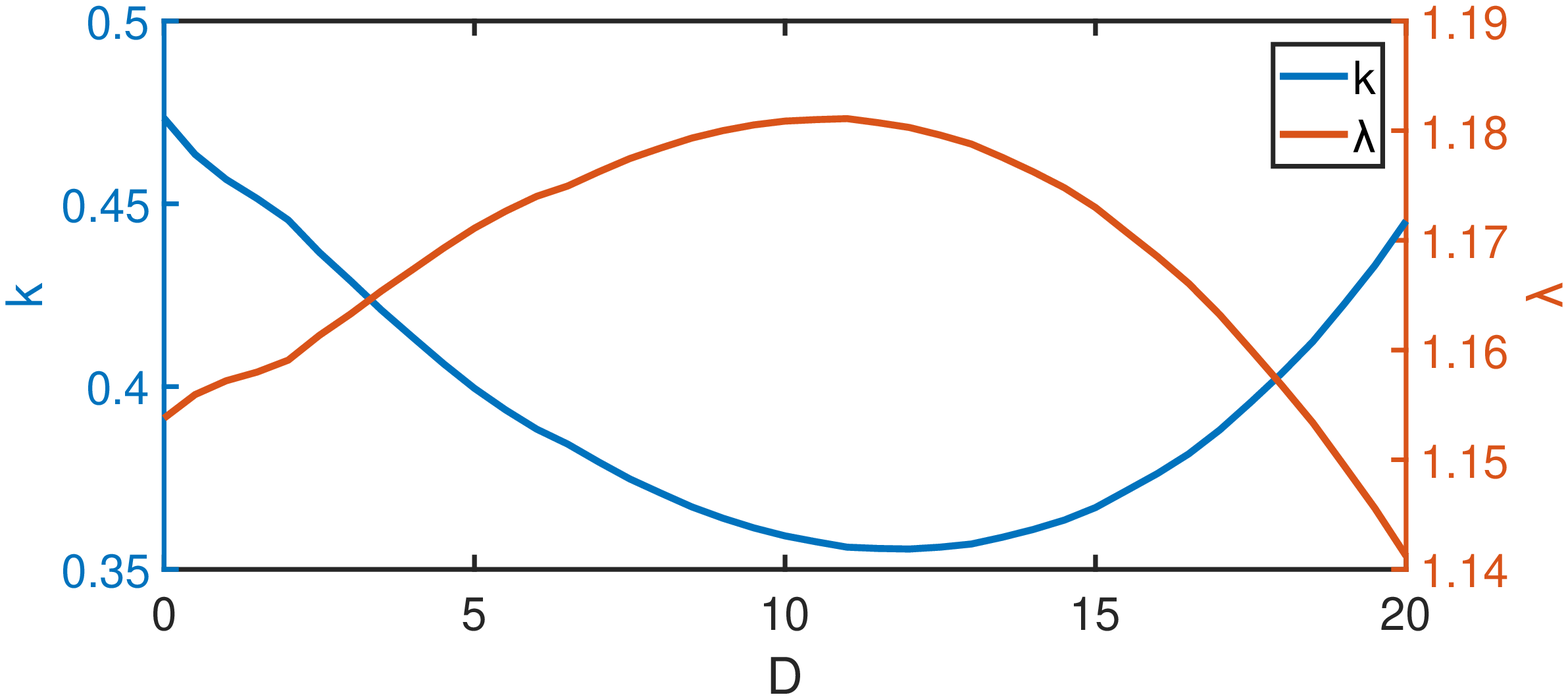}\vspace{-10pt}
	\caption{\label{figdensigy} The impact of different $ D $ parameters on $ k $ and $\lambda$. When $ D=0 $, the potential corresponds to an quartic potential with $ k=0.47 $ and $ \lambda=1.15 $, which is in agreement with Ref.\cite{bastin2017ion}.}
	\label{kD}
\end{figure}

In order to find the critical value $\beta _c$ at which the zigzag transition occurs, we search when the minimum eigenvalue $A=0$ as $\beta$ decreases, thus obtaining the critical point $\beta_c$ for the phase transition as a function of given $N$. Then we use a well-known simple power form relationship: \upcite{schiffer1993phase,dubin1993theory,hughes1998alamos,enzer2000observation,morigi2004dynamics,lin2009large}
\begin{equation}
	\beta_c=kN^\lambda	
\end{equation}
for fitting, since this relationship exist in both purely quadratic and purely quartic potential. During the calculation, the location of each particle $z_i$ is determined from the continuous model by using the reciprocal function of the integration of the density function. In Fig.\ref{kD}, we demonstrate the impact of $D$ parameters on $k$ and $\lambda$. 

\begin{figure}[htbp]
	\centering
	\subfigure{\includegraphics[width=3.3in]{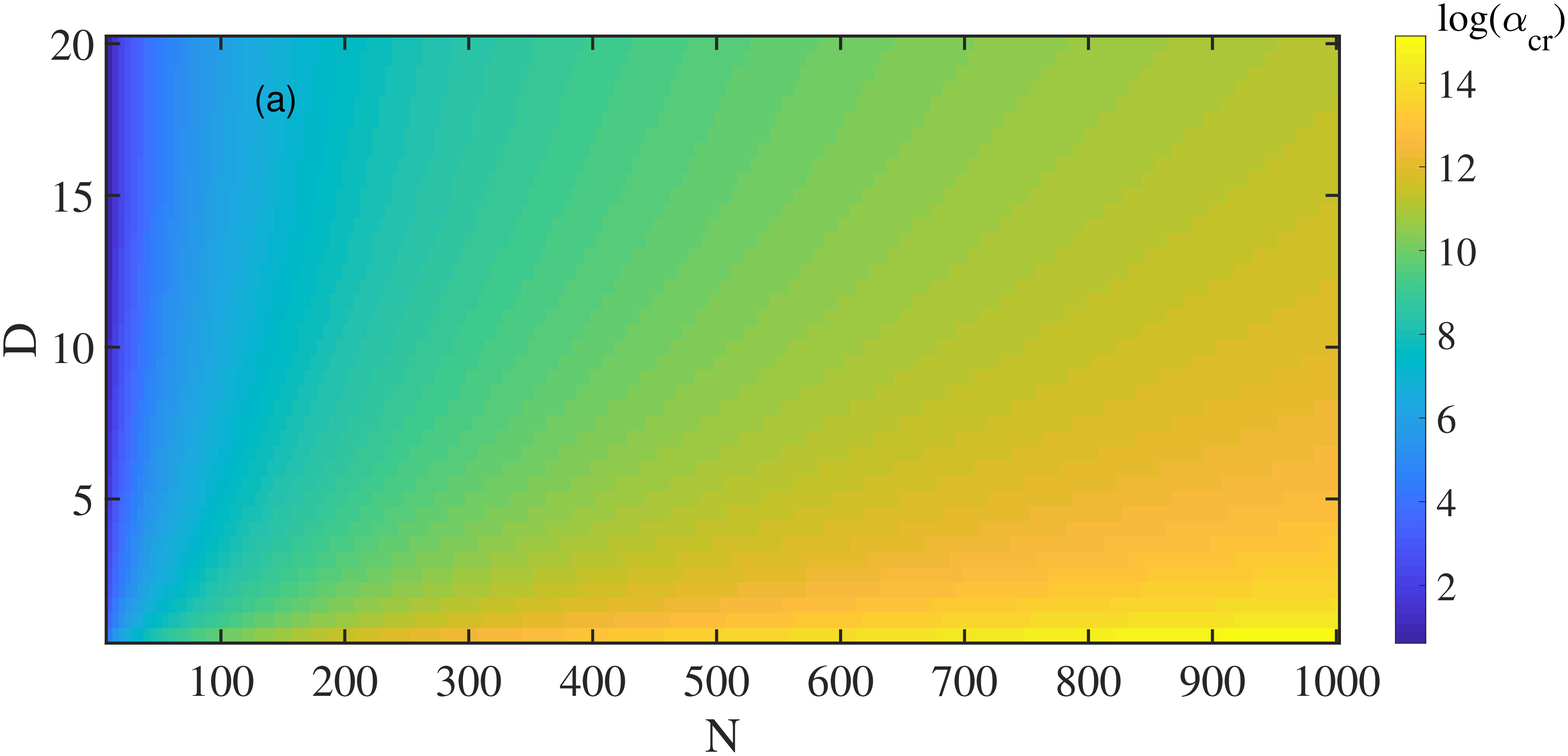}}\vspace{-10pt}
	\subfigure{\includegraphics[width=3.3in]{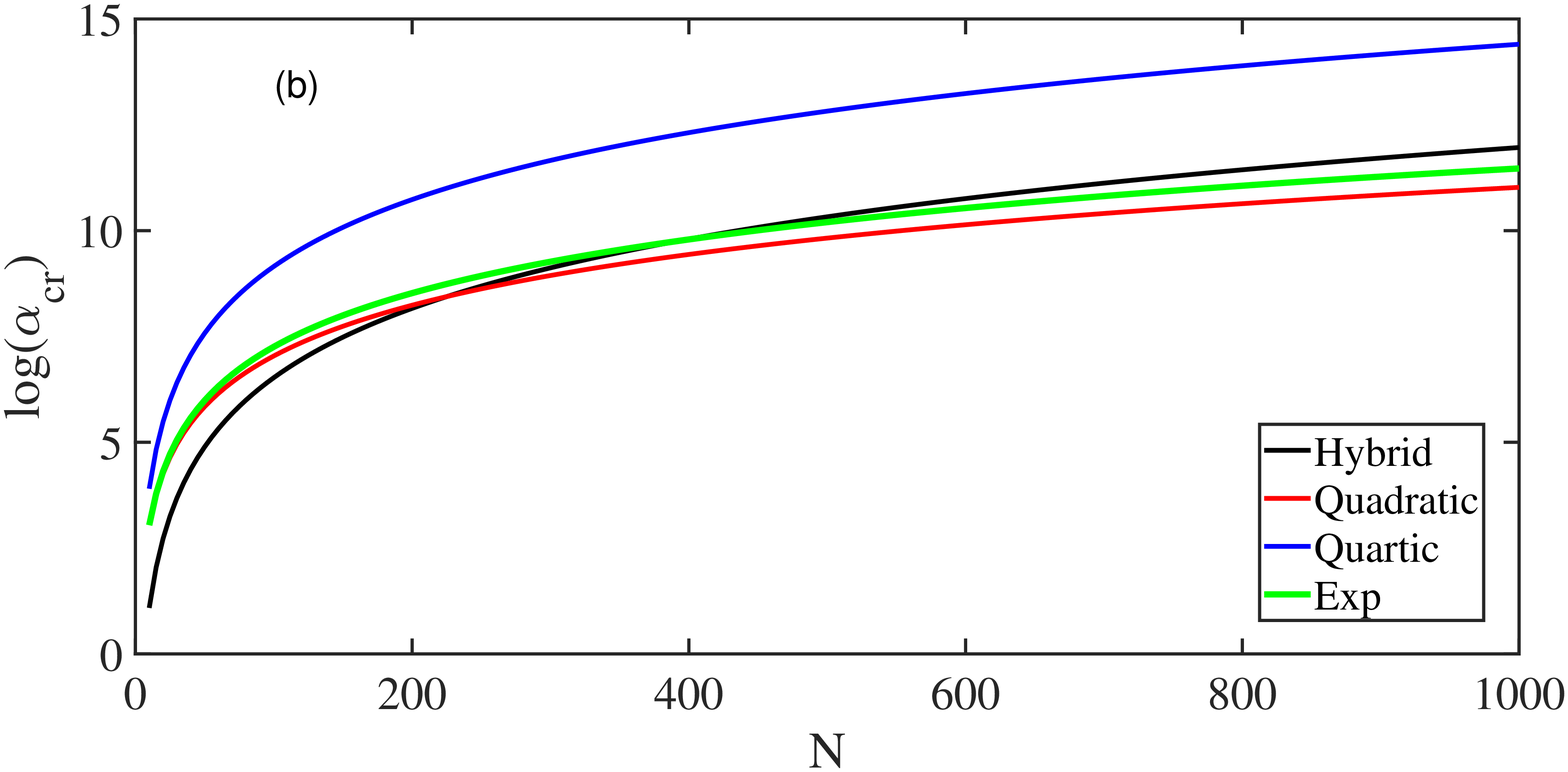}}\vspace{-10pt}
	\subfigure{\includegraphics[width=3.3in]{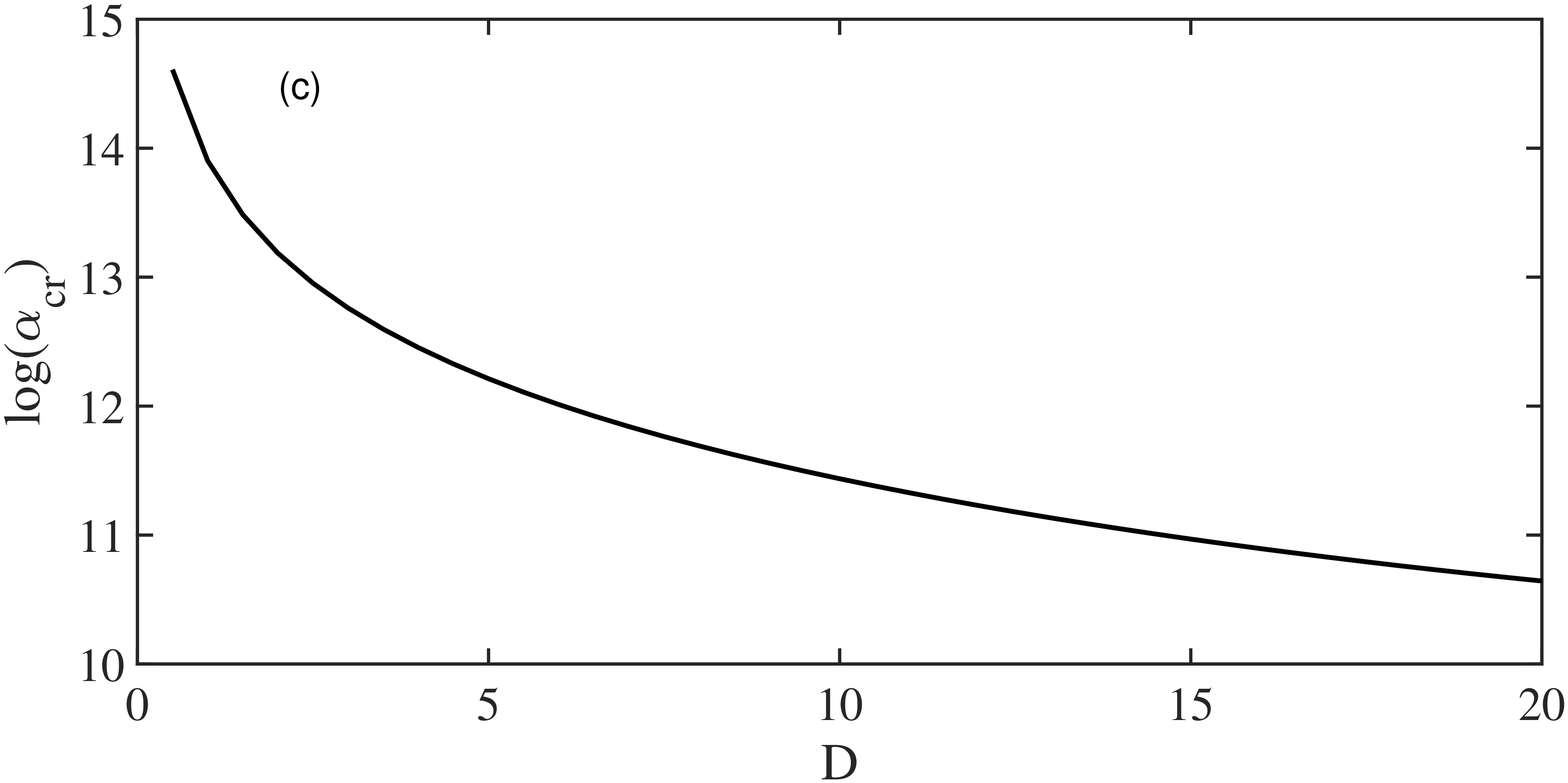}}\vspace{-10pt}
	\caption{\label{figdensigy} (a)The phase diagram showing the dependence of critical anisotropy $\alpha_{cr}$ on both $ D $ and $N$ parameter. This critical anisotropy $\alpha_{cr}$ determins when the phase transition from linear chain to zigzag phase occurs. (b)When $ D=10 $, the critical anisotropy $\alpha_{cr}$ as a function of $N$ for the hybrid potential(solid black line), pure quadratic potential(solid red line), \upcite{schiffer1993phase,dubin1993theory,hughes1998alamos,enzer2000observation}and pure quartic potential(solid blue line), \upcite{bastin2017ion}as well as the function(solid green line) obtained from experimental data.\upcite{enzer2000observation} (c) When $N=800$, the critical anisotropy $\alpha_{cr}$ as a function of $D$ for the hybrid potential.}
	\label{phase}
\end{figure}

With this relationship, we further map out the phase diagram for this phase transition, which is shown in Fig.\ref{phase}. In Fig.\ref{phase}(a), the dependences of critical anisotropy $\alpha_{cr}$ on the $ D $ and $N$ parameter are shown. Clearly, $\alpha_{cr}$ is much larger at large $N$ and small $D$. While it is much smaller at small $N$ and large $D$. Fig.\ref{phase}(b) shows $\alpha_{cr}$ as a function of $N$ when $D = 10$  for the hybrid potential(solid black line), pure quadratic potential(solid red line),\upcite{schiffer1993phase,dubin1993theory,hughes1998alamos,enzer2000observation} and pure quartic potential(solid blue line).\upcite{bastin2017ion} The function obtained from experimental data \cite{enzer2000observation} is also shown in solid green line. Fig.\ref{phase}(c) shows $\alpha_{cr}$ as a function of $D$ when $N = 800$  for the hybrid potential. As it is shown, $\alpha_{cr}$ increases with $N$ in a power form, more importantly, $\alpha_{cr}$ in our cases lies between that of pure quadratic and quartic potential, and reach closely to the value inferred from experimental data \upcite{enzer2000observation} for large $N$. Such a good agreements verifies the accuracy and the uo niversality of our proposal. 

\section{Conclusion}

We have studied the properties of the one-dimensional ion crystal in a hybrid potential. Both the density distribution and the optimal parameter $D$ are attained by both the discrete model and the continuous model. Large homogeneous ion chain with ion number more than 800 have been achieved, which has never been realized before. In principle, the number of ions can be increased to tens of thousands, even millions in such hybrid potential, if the trapping space allowed. Based on the result from continuous model, we calculate the critical condition for the zigzag transition. When the zigzag transition occurs, the $\beta_c$ can be controlled by adjusting the parameter $ D $, as Fig.\ref{kD} illustrates. 

Such potential can accommodate more ions uniformly distrubuted than both quadartic potential and quartic potential, avoiding zigzag transitions at the center of the chain. Additionally, it can prevent the ions in the center coming too close, thus reducing cross-talk in ion state detection and enabling single ion manipulation with focused laser beams. Particularly, it offers unique advantages in performing coherent operations on large ion chain.\upcite{lin2009large,lin2016sympathetic} It can also be used to investigate the effects of the quartic term in the hybrid potential on the normal mode structure.\upcite{home2011normal,johanning2016isospaced} As the spin-spin interaction approximately scales to the second power of the distance, a homogenious ion chain can be used for quantum simulation of Ising models.\upcite{porras2004effective} 

Moreover, the large ratio $\alpha_{cr}$ of the radial potential and the axial potential provided by the homogenious ion chain, given the required hybrid potential, would enable significant improvement on the gate fidelity, since the thermal ion motion can be reduced by a factor ranging from $\alpha_{cr}^2$ to  $\alpha_{cr}^3$.\upcite{zhu2006trapped} Our proposal is also a promising route for scaling up towards a large-scale quantum computer in a single harmonic trap, and could potentially open up a new avenue for the study of large-scale quantum information processing. 

\section{Acknowledgement}

This work was supported by the Natural Science Foundation of Guangdong Province of China (Grants No. 2017A030310452, 2020A1515010864, 2020A1515011159), the National Natural Science Foundation of China (Grants No.11904423, 11974434), and Fundamental Research Funds for the Central Universities of Education of China(Grants No.17lgpy27, 191gpy276). 



\end{document}